# Moiré-induced transport in CVD-based small-angle twisted bilayer graphene

Giulia Piccinini[1,2], Vaidotas Mišeikis[2,3], Pietro Novelli[4], Kenji Watanabe[5], Takashi Taniguchi[6], Marco Polini[7,3], Camilla Coletti[2,3,*], Sergio Pezzini[8,**]

[1]*NEST, Scuola Normale Superiore, Piazza San Silvestro 12, 56127 Pisa, Italy*
[2]*Center for Nanotechnology Innovation @NEST, Istituto Italiano di Tecnologia, Piazza San Silvestro 12, 56127 Pisa, Italy*
[3]*Graphene Labs, Istituto Italiano di Tecnologia, Via Morego 30, 16163 Genova, Italy*
[4]*Istituto Italiano di Tecnologia, Via Melen 83, 16152 Genova, Italy*
[5]*Research Center for Functional Materials, National Institute for Materials Science, 1-1 Namiki, Tsukuba, 305-0044, Japan*
[6]*International Center for Materials Nanoarchitectonics, National Institute for Materials Science, 1-1 Namiki, Tsukuba, 305-0044, Japan*
[7]*Dipartimento di Fisica, Università di Pisa, Largo Bruno Pontecorvo 3, 56127 Pisa, Italy*
[8]*NEST, Istituto Nanoscienze-CNR and Scuola Normale Superiore, Piazza San Silvestro 12, 56127 Pisa, Italy*

**Abstract**

To realize the applicative potential of 2D twistronic devices, scalable synthesis and assembly techniques need to meet stringent requirements in terms of interface cleanness and twist-angle homogeneity. Here, we show that small-angle twisted bilayer graphene assembled from separated CVD-grown graphene single-crystals can ensure high-quality transport properties, determined by a device-scale-uniform moiré potential. Via low-temperature dual-gated magnetotransport we demonstrate the hallmarks of a 2.4°-twisted superlattice, including tunable regimes of interlayer coupling, reduced Fermi velocity, large interlayer capacitance, and density-independent Brown-Zak oscillations. The observation of these moiré-induced electrical transport features establishes CVD-based twisted bilayer graphene as an alternative to `tear-and-stack' exfoliated flakes for fundamental studies, while serving as a proof-of-concept for future large-scale assembly.

**Main Text**

Twisted 2D materials provide an extraordinarily rich platform for engineering emergent electronic [1, 2], magnetic [3] and optical [4] properties. The van der Waals (vdW) stacking techniques [5, 6] – not applicable to traditional low-dimensional condensed-matter systems [7] – are especially boosting this research field, allowing the realization of complex moiré structures involving multiple precisely aligned atomic layers [8–10]. As *twistronics* – the understanding and control of the moiré-induced behaviours – rapidly advances [11–13], novel perspectives of technological application arise [14]. For instance, the tantalizing superconducting phase of magic-angle (MA) twisted bilayer graphene (TBG) [1] has already been exploited for the fabrication of broadband photodetectors [15], as well as gate-defined monolithic Josephson junctions [16–18] and quantum interference devices [19]. However, while for stand-alone 2D materials technological integration appears increasingly viable thanks to key advancements in the synthesis methods [20] (such as chemical vapour deposition (CVD) of high-mobility single-layer graphene (SLG) [21–24]), in the case of TBG, further challenges have to be addressed. Ideally, application-oriented TBG devices should simultaneously offer: (i) a deterministically-selectable small-angle (SA) twisting, (ii) a device-scale uniform twist angle, (iii) an atomically-clean interlayer enabling the formation of a moiré potential. In TBG, strong modifications in the electronic bands arise only for SA twisting ($\theta < 5°$) [25, 26], while the physics of two decoupled layers is reached asymptotically at larger twist angles [27–29]. SA twisting was observed in CVD-grown graphene films studied by scanning probe microscopy [30]. However, due to its polycrystalline nature and random grain orientations, this system is unsuitable for spatially-averaging probes such as electrical transport. CVD-grown graphene single crystals, compatible with fabrication of high-quality devices, can incorporate TBG domains with uniform twisting [31–34]. Nonetheless, the twist angle preferentially locks to $0°$ (Bernal stacking) or $30°$ due to interactions with the growth substrate [31].

Recent developments in the synthesis process [35] have allowed to obtain a fraction of intermediate twist-angles (down to $\sim 3°$), higher than in previous studies [36] but lacking however deterministic control, as well as moiré transport signatures. To overcome this issue, one can employ a *hybrid* approach by stacking two CVD-grown SLG to form TBG, obtaining either large or SA twisting, as demonstrated by photoemission [37, 38, 39] and scanning probe experiments [40, 41], respectively. Although permitting high rotational accuracy, in analogy to exfoliated flakes [6], sequentially-stacked CVD-grown graphene layers tend to damage and accumulate contaminants at their interface [42]. As a consequence, transport experiments on CVD-based TBG with moiré effects are (to the best of our knowledge) unreported and, therefore, a conclusive demonstration of TBG realizing the preliminary scalability conditions outlined above is lacking.

In this work, we fill this gap by introducing SA-TBG samples obtained by hBN-mediated stacking of isolated SLG crystals grown by CVD on a single Cu grain. The growth-determined crystallographic alignment of the SLG crystals [43] enables deterministic control on the twist angle at the vdW assembly stage. The interface cleanness and twist-angle uniformity are unambiguously supported by the observation of high-quality quantum transport features specific to TBG with a twist angle of $\sim 2.4°$. By these means, we demonstrate the first moiré device based on CVD-grown crystals and set a cornerstone toward the application of 2D materials twistronics.

In Figure 1a we present the vdW assembly sequence developed for CVD-based SA-TBG. As pick-up medium we employ a poly(bisphenol A carbonate)(PC) film deposited onto a few-mm thick polydimethylsiloxane (PDMS) block, supported by a glass slide [44], that we control using a home-built transfer setup [45]. We start from an array of SLG crystals grown via CVD on Cu (see the SI file for details) and subsequently transferred to SiO$_2$/Si using a polymer-assisted technique, as described in Refs. [43, 45]. We select two graphene crystals from the array, making sure that they were synthesized on the same Cu grain and, therefore, that they share the same crystallographic orientation, as demonstrated in Ref. [43] (Figure 1b). The use of two separated crystals extends the standard method for preparing SA-TBG samples for transport studies, which proceeds by stacking two portions of the *same* SLG flake [6]. Once the two crystals are selected, we adopt the procedure described in Ref. [44] to pick up the first graphene crystal from SiO$_2$ using an hBN flake (10–50 nm thick). We then use the goniometer stage holding the sample with graphene on SiO$_2$ (shown in SI Figure S1) to rotate the graphene array by an arbitrary angle $\theta$, which is affected by an instrumental error of $\sim 0.01°$. The twist angle $\theta$ determines the expected periodicity $\lambda$ of the moiré pattern (Figure 1c), according to $\lambda = \frac{a}{2\sin(\frac{\theta}{2})}$, where $a \simeq 0.246\ nm$ is the SLG lattice constant. Thereafter, we approach and pick up the second graphene crystal, and a second hBN flake, completing the encapsulation. The temperature of the setup is kept at 40-60°C during all these steps, consistently ensuring the complete pick-up of the graphene regions approached by the hBN. Finally, the stack is released onto a SiO$_2$/Si substrate by melting the PC film at 160-170°C, favouring cleaning of the vdW interfaces [44]. After the assembly, we nevertheless observe blisters where contaminants aggregate (Figure 1d, inset), which limit the lateral dimension of flat areas suitable for device processing (typically few micron-wide). Figure 1d shows the Raman spectrum of the assembled TBG, compared to that of an hBN-encapsulated SLG. The two spectra differ in several features. The large 2D/G intensity ratio characteristic of SLG ($\sim 10$) dramatically drops in TBG ($\sim 1$). In addition, the 2D peak width strongly increases, from $\sim 17\ cm^{-1}$ to $\sim 54\ cm^{-1}$. At a closer inspection, the 2D peak of TBG reveals a multicomponent structure [46–48], with two broad sub-peaks located at $\sim 2675\ cm^{-1}$ and $\sim 2700\ cm^{-1}$. Overall, we observe striking similarities with the Raman spectrum at $\sim 2.6°$-twisting reported in Ref. [46], in accordance with the angle $\theta = 2.5°$ set during the vdW assembly. The assembly of a second sample with the same target angle, showing analogous Raman response, is presented in SI. Raman data from a third sample with sub-MA twisting are shown in the SI file.

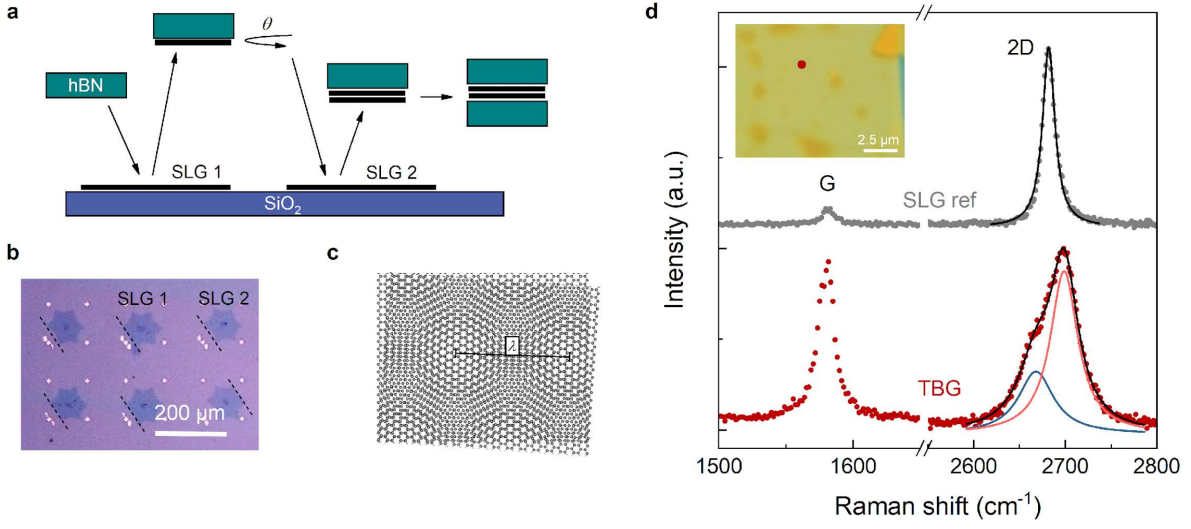

**Figure 1.** (a) Schematics of the dry pick-up process with stacking of separated CVD-grown graphene crystals. (b) Optical microscopy image of CVD SLG crystals on SiO$_2$/Si. The dashed lines indicate their crystallographic alignment. (c) The $\theta$-rotated graphene sheets form a moiré pattern with periodicity $\lambda$. (d) Representative Raman spectrum of TBG (dark red), compared to a SLG reference (grey). The light red and blue lines are the two Lorentzian components of the TBG 2D peak. Inset: optical microscopy image of hBN-encapsulated SA-TBG. The dark red spot indicates the point where the TBG spectrum in the main panel is acquired.

The target angle $\theta$ is chosen to fall in the intermediate twist-angle range where the Fermi velocity ($v_F$) is reduced with respect to that of SLG [25], while the interlayer coupling can be varied from weak to strong by experimentally-available gate voltages. Such tunability was demonstrated by transport experiments on devices obtained by `tear-and-stack' exfoliated flakes in Refs. [47–51], which serve as a guideline for our investigation of CVD-based SA-TBG.

In Figure 2 we show low-temperature (magneto)transport data on a dual-gated device fabricated from the SA-TBG sample (see SI for details on the processing). The dual-gated configuration (Figure 2b) is essential in multilayer graphene devices, as it allows independent tuning of the total carrier density ($n_{tot}$, determined by the sum of the gate potentials) and its distribution among the layers via the so-called displacement filed ($D$, determined by the difference of the gate potentials). However, this holds true as long as the interlayer coupling is small enough as to keep the layers' Dirac cones independent [27–29, 34], while in the strong coupling regime $D$ has no major effect [52]. By applying a perpendicular magnetic field ($B = 3\,T$ in Figure 2a) we observe a pattern of crossings in the derivative of the Hall conductivity $\sigma_{xy}$ with respect to the voltage applied via the top gate ($d\sigma_{xy}/dV_{tg}$), corresponding to alternating interlayer quantum Hall states ($d\sigma_{xy}/dV_{tg} = 0$) and layer-resolved Landau levels (LLs, $d\sigma_{xy}/dV_{tg} \neq 0$) [34]. This pattern can be modelled by considering the screening properties of two superimposed SLG subject to the top and bottom gate potentials— $V_{tg}$ and $V_{bg}$, respectively

—and coupled via an interlayer capacitance $C_{gg}$ [27, 29] (complete details on the electrostatic model employed can be found in Ref. [53]). Importantly, the exact gate dependence of the LLs is sensitive to both the carrier density $n$ and Fermi energy $E_F$ in the individual layers, which for Dirac fermions are related according to $E_F = \hbar v_F \sqrt{\pi n}$. Using $C_{gg}$ and $v_F$ as free parameters, we simulate the LLs trajectories, and make them converge to the experimental pattern of crossings. The results are shown as orange and red dotted lines in Figure 2a, for the upper and lower layers LLs, respectively. From this procedure, we can estimate a Fermi velocity $v_F = (0.47 \pm 0.02) \times 10^6 \, m/s$ and an interlayer capacitance $C_{gg} = (17.5 \pm 1.0) \times 10^{-6} \, F/cm^2$.

The suppression of $v_F$ with respect to SLG is a well-known feature of SA-TBG [25, 30]. Band structure calculations based on Bistritzer-MacDonald-type Hamiltonians [25, 54, 55] allow to estimate the corresponding twist angle to be $\sim 2.4°$ (Figure 2c).

Concerning the interlayer capacitance and in agreement with Ref. [48], our estimate is twice as large with respect to the accepted value of $C_{gg}$ for large-angle TBG [27, 29]. If one insists in using a classical-type formula for $C_{gg}$, i.e. $C_{gg} = \varepsilon_0 \varepsilon_r / d_{eff}$, with $d_{eff}$ a suitable effective inter-layer distance, this finding could be interpreted in terms of a smaller *effective* interlayer spacing, signaling the increased coupling in this twist-angle range (eventually, toward MA such effective separation vanishes, leading to a complete suppression of the LLs crossings [52]). A less naïve approach should rely on analyzing microscopically all the non-classical contributions to $C_{gg}$ by using the profound relationship that exists between the ground-state energy of a double-layer system and linear response functions [56]. This has been recently done for example in Ref. [57], but no explicit calculations have been reported by the authors for TBG.

The crossing pattern in Figure 2a is abruptly interrupted in the vicinity of the upper right and lower left corners of the $V_{tg} - V_{bg}$ map, i.e. at high total carrier density ($n_{tot} > 5.88 \times 10^{12} \, cm^{-2}$; $n_{tot}$ being the sum of the carrier densities in the two layers obtained from the electrostatic modelling). The Hall conductivity $\sigma_{xy}$, plotted in Figure 2d, shows that the upper right (lower left) region corresponds to a transition from large electron (hole) density to large hole (electron) density. This change contrasts with the low-density switch at the charge-neutrality point (CNP, central diagonal), and it is characteristic of van Hove singularities (vHs) in the density of states, corresponding to the transition from layer-independent massless electrons (holes) to layer-coupled massive holes (electrons) [47–51]. In Figure 2e, we show the zero-field longitudinal conductivity as a function of the gate potentials in the vicinity of the sample CNP (black-highlighted area in Figure 2d). In this zoomed plot we can observe different regions of interlayer charge configuration, controlled by a splitting of the CNPs of the individual layers (similar data for large-angle TBG have been reported in Refs. [53, 58]). The boundaries of these regions are perfectly reproduced by the neutrality conditions for the two layers (orange and red dotted lines), computed according to the extracted $v_F$ and $C_{gg}$, confirming the estimate obtained in the perpendicular magnetic field (in the SI file we show how the CNPs trajectories vary as functions of $v_F$ and $C_{gg}$). The observation of the CNPs splitting substantiates the reduction of the peak resistance at large displacement fields observed in previous experiments [47, 50, 51] which is due to coexisting charges of opposite sign in the two layers with relatively small concentration ($< 10^{11}\ cm^{-2}$).

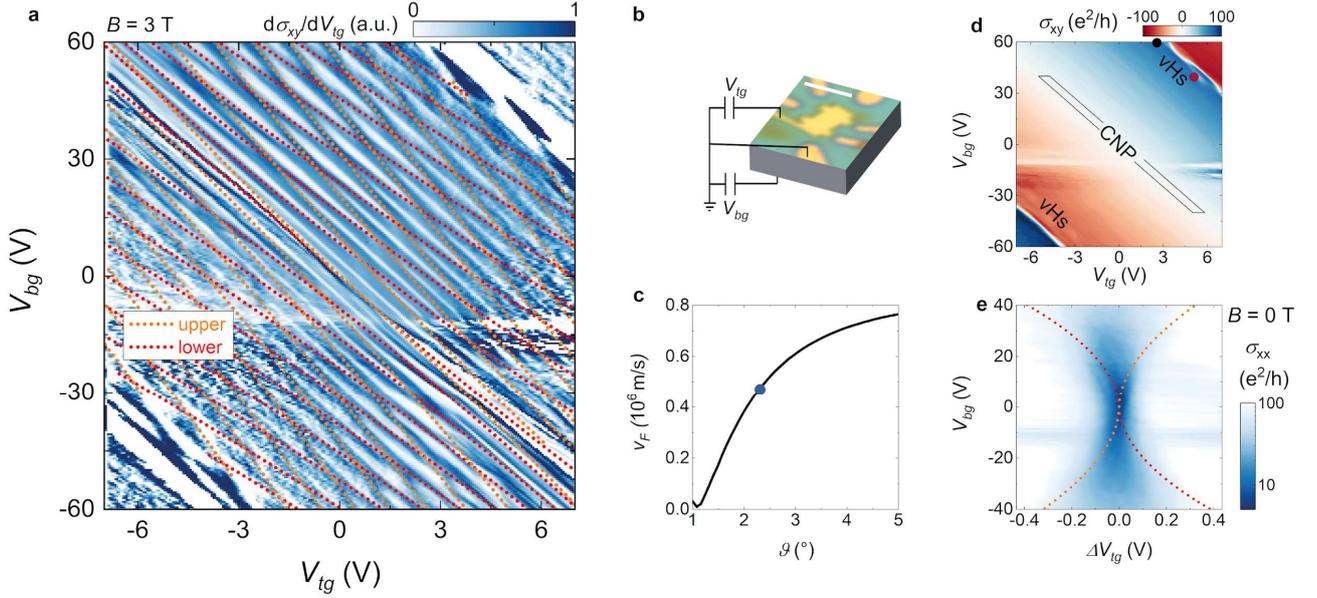

**Figure 2.** (a) First derivative of the Hall conductivity as a function of top and back-gate voltages, measured for a fixed value of the applied perpendicular magnetic field ($B = 3\ T$). The dotted orange (red) lines are the calculated positions of Landau levels from the upper (lower) graphene layers, employing $v_F = 0.47 \times 10^6\ m/s$ and $C_{gg} = 17.5 \times 10^{-6}\ F/cm^2$. (b) Schematics of the gating configuration. The optical microscopy image of the device is taken before the final etching step; the scale bar is 2.5 $\mu m$. (c) Fermi velocity of TBG as a function of the twist angle, calculated according to the theory described in Refs. [54, 55] and references therein. Results in this figure have been obtained by setting $u_0 = 79.7\ meV$ and $u_1 = 97.5\ meV$, where $u_0$ and $u_1$ are the intra- and inter-sublattice inter-layer tunneling amplitudes, respectively. The blue circle corresponds to the $v_F$ value estimated for our device. (d) Hall conductivity as a function of the gate voltages (same ranges and magnetic field as in (a)). The sign changes in $\sigma_{xy}$ correspond to the sample CNP and the two vHs. The black rectangle indicates the gate range considered in panel (e), the black and dark red dots are the gate values used for the measurements in Figure 4. (e) Zero-field longitudinal conductivity (Log scale) as a function of top-gate voltage relative to the sample CNP, and back-gate voltage. The dotted orange (red) line is the calculated charge neutrality point for the upper (lower) layer. All the data in this figure have been acquired at $T = 4.2\ K$.

In Figure 3a we present the longitudinal resistance of the device, measured as a function of $V_{tg}$ and $B$, at two fixed values of $V_{bg}$ (-60 V and +60 V, in the left and right panels, respectively, corresponding to the upper and lower limits of the gate map in Figure 2a). This configuration is chosen to span the largest possible density range, while keeping $D$ finite. A fast Fourier transform (FFT) of these data, giving the frequency spectrum of the $1/B$- periodic components of the resistance [47], is shown in Figure 3b to ease the interpretation of the complicated pattern of experimental data reported in Figure 3a. The color map in Fig. 3b represents the normalized amplitude of the FFT plotted as a function of the total carrier density, and the frequency $B_F$, which is proportional to the extremal area of the Fermi surface perpendicular to the magnetic field. In the central part of the magneto-resistance data, close to the CNP, we observe two superimposed Landau fans, which can be attributed to the upper and lower layers' Dirac cones centred at the $K_s$ and $K_s'$ points in the superlattice Brillouin zone (see band structure calculations in Figure 3c, inset). Due to the finite displacement field, the two layers have different carrier concentrations and, therefore, the fans repeatedly cross each other. The corresponding frequencies in the FFT are well described by $B_F^{layer} = \frac{h}{4e}|n_{layer}|$ (orange and red dotted lines in Fig. 3b, for the upper and lower layer, respectively), where the carrier concentration $n_{layer}$ in each layer can be calculated by using the $v_F$ and $C_{gg}$ values extracted previously, and the factor 4 accounts for the spin and valley degeneracies. The sum of the two components evolves as $B_F^{sum} = \frac{h}{4e}|n_{tot}|$ (dark red dotted line).

At the largest density reached in the experiment (left-most part in the left panel, right-most part in the right one), we observe LL fans with opposite dispersions with respect to the central ones, in accordance with the change in sign of the charge carriers detected in the Hall conductivity (Figure 2d). These features emerge due to progressive filling of the moiré band at $\Gamma_s$ (Figure 3c, inset), which is completed at $n_{tot} \equiv n_s = \frac{8}{\sqrt{3}\lambda^2}$. In Figure 3b we show that the frequency components corresponding to the $\Gamma_s$ fans evolve as $B_F^{\Gamma_s} = \frac{h}{4e}|n_{tot} - n_s|$ (dark cyan dotted lines, where $n_s = \pm 13.3 \times 10^{12}\ cm^{-2}$ for the right and left panel, respectively), indicating a single four-fold degenerate Fermi surface, in agreement with Ref. [47]. In Figure 3c we show that the corresponding fan of quantized states, calculated using a zero Berry phase [49], matches the resistance oscillations in panel a.

Close to the previously identified vHs ($V_{tg} \sim -4V$ and $+4V$, in the left and right panel, respectively) we observe two funnelling structures with large longitudinal resistance, associated to the coexistence of carries with opposite sign. Here, the oppositely dispersing fans of Landau levels coalesce, as expected from theoretical calculations in our twist-angle range [26, 59]. Notably, we observe a series of horizontal strikes superimposed to the intersecting fans, which signal a density-independent oscillation of the resistance. The corresponding frequency is equal to $137\ T$ (magenta dotted line in Figure 3b). Density-independent oscillations were discovered in graphene-hBN superlattices [60–62], and attributed to the periodic creation of so-called Brown-Zak (B-Z) particles moving along straight trajectories in finite magnetic field [61]. The characteristic frequency of this phenomenon allows a highly precise estimate of the moiré periodicity according to $B_F^{BZ} = \frac{h}{e}\frac{2}{\sqrt{3}\lambda^2}$. Considering the average position of our FFT peak, we obtain a twist angle $\theta = (2.39 \pm 0.01)°$. Finally, in Figure 3d we show the Hall conductivity in the vicinity of the hole-side vHs, as a function of $1/B$. In accordance to the B-Z periodicity, we observe sign changes at commensurate values of flux quanta per superlattice unit cell $\phi/\phi_0 = 1/q$ (where $\phi_0 = h/e$ is the flux quantum, and $q$ is an integer). In addition, towards the highest magnetic fields, we observe a non-monotonic behavior as a function of both magnetic field and carrier density, a hallmark of the Hofstadter's butterfly [63–65]. The appearance of these features coincides with the transition from the semiclassical regime (well-defined electron and hole-like oscillations) to the fractal regime, which is expected when the magnetic length ($l_B \sim \frac{25\ nm}{\sqrt{B[T]}}$) becomes comparable to the superlattice periodicity $\lambda = 5.9\ nm$ [26].

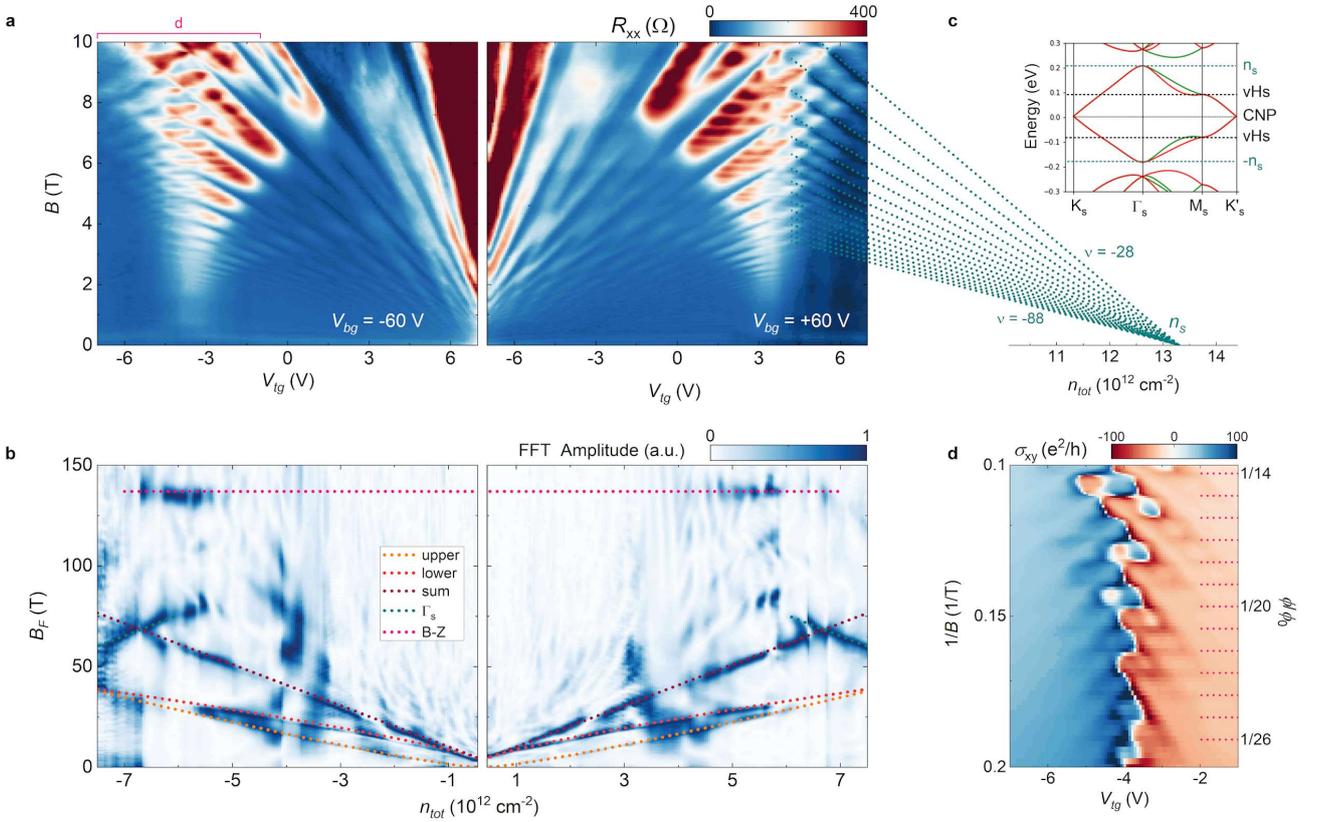

**Figure 3.** (a) Longitudinal resistance measured as a function of $V_{tg}$ and $B$, at $V_{bg} = -60\ V$ (left panel, $T = 2.5\ K$) and $V_{bg} = +60\ V$ (right panel, $T = 4.2\ K$). (b) Normalized FFT amplitude of the data in panel (a), as a function of the total charge density and of the oscillation frequency $B_F$. (c) Fan of quantized states originating from the $\Gamma_s$ point. Inset: band structure calculations for TBG with $\theta = 2.4°$, based on Refs. [25, 54, 55]. The same intra- and inter-sublattice inter-layer tunneling amplitudes of Figure 2c are used. Hartree self-consistent corrections do not yield significant changes with respect to single-particle calculations because the twist angle considered in this work is sufficiently larger that the MA. (d) Hall conductivity in the vicinity of the hole-side vHs, as a function of $V_{tg}$ and $1/B$ (left axis). The right axis scale shows the number of flux quanta per superlattice unit cell, i.e. $\phi/\phi_0$.

A distinctive feature of the B-Z oscillations is their resilience to the thermal energy, which allows their observation up to boiling-water temperature [60]. In Figure 4 we present resistance data acquired at $T = 35\ K$, where the standard Shubnikov–De Haas oscillations are strongly suppressed and the B-Z oscillations become more apparent [50]. We show two curves taken in the vicinity of the electron-side vHs, at $D = 0$ and $D > 0$ (dark red and black curves, respectively; the gate values are indicated by markers in Figure 2d). We observe a dominant fast oscillation corresponding to the B-Z frequency ($B_F = 137\ T$, see FFT spectra in the inset), whose amplitude and phase are unaffected by $D$ (in addition to $n_{tot}$, as already shown in Figure 3b). This contrasts with the slowly varying background ($B_F \sim 30\ T$), attributed to the $K_s$-$K_s'$ Shubnikov–De Haas oscillations, whose phase reverts as a function of $D$ as the charge distribution in the two layers is modified. Recently-discovered $D$-dependent high-temperature oscillations from inter-minivalley scattering [66] are not observable in our current set of data.

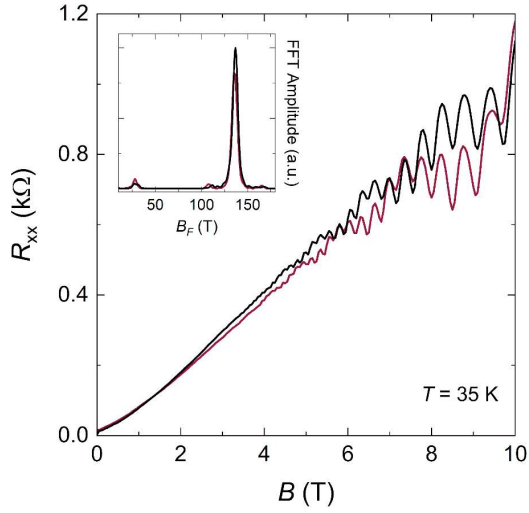

**Figure 4.** Longitudinal resistance as a function of $B$, measured at $T = 35\ K$ in the vicinity of the electron-side vHs, at $D = 0$ (dark red curve) and $D > 0$ (black curve); the gate values are indicated by the dark red and black circles in Figure 2d. Inset: FFT spectra of the oscillatory resistance from the curves in the main panel.

The experimental observation of this collection of moiré-induced transport features necessarily implies the presence of a superlattice with uniform twist angle (within cent-of-degree accuracy) over the device area within the voltage probes ($\sim 2\ \mu m^2$). While local techniques have been successfully applied before to CVD-based SA-TBG [40], transport studies are not available in the literature, to the best of our knowledge. The realization of device-scale moiré effects using CVD-grown crystals is of high relevance for different potential applications. In particular, TBG can be used for ultrafast, highly sensitive and selective photodetectors [15, 67]. Moreover, moiré patterns in SA-TBG provide confined conducting channels that can be used for the directed propagation of surface plasmons [68] or for the study of moiré plasmons [54, 55, 69].

In principle, our assembly approach could be up-scaled by employing multiple crystals from the same array simultaneously (i.e. within the same pick-rotate-and-stack process). Nonetheless, two main limiting factors to scalability of CVD-based SA-TBG should be considered. First, the presence of blisters due to incomplete interface cleaning currently constrain the device dimensions: this could be mitigated by using dry polymer-free techniques for CVD graphene transfer, such as in Ref. [21]. Secondly, the requirement of hBN flakes – acting both as pick-up carrier and high-quality electrostatic environment for TBG – which are limited to the lateral size currently yielded by micromechanical exfoliation (typically up to ~100 μm).

In addition, while a path toward twisted *N*-layer graphene devices appears to be traced by recent results on flake-based quadrilayers and pentalayers [70, 71], a crucial experimental bottleneck arises. Exfoliated graphene flakes have limited lateral dimensions (up to ~100 μm), which impede the realization of thick angle-controlled stacks with areas compatible with device fabrication. Since our CVD matrixes retain a single crystallographic orientation over mm-sized areas, stacking of graphene layers with N>5 and device-compatible size could be pursued using the technique introduced here.

In conclusion, we demonstrated the first SA-TBG high-quality moiré device based on CVD-grown crystals. The use of aligned graphene crystals from CVD-grown arrays, together with the manual stacking approach, allows deterministically-selectable twist angles. The existence of a moiré potential with uniform periodicity on a device-scale area is confirmed by the observation of density-independent Brown-Zak oscillations, which coexist with multiple Landau fans at low temperature, and survive up to tens of Kelvin. Overall, our results establish a novel tool for future developments of 2D materials twistronics and related technology.

**Associated Content**

*Supporting Information:* Details on the experimental methods for CVD growth, Raman spectroscopy, device fabrication, and low-temperature magnetotransport; assembly of a second SA-TBG sample; Raman and transport data on CVD-based bilayer graphene with sub-MA twisting; dependence of the CNPs splitting on Fermi velocity and interlayer capacitance; Figures S1–S4.

**Author Information**

*Corresponding Authors:*
* camilla.coletti@iit.it
** sergio.pezzini@nano.cnr.it

*Notes:*

The authors declare no competing financial interest.

**Acknowledgments**

We thank F. Rossella for technical support during the low temperature experiments. Growth of hexagonal boron nitride crystals was supported by the Elemental Strategy Initiative conducted by the MEXT, Japan, Grant Number JPMXP0112101001, JSPS KAKENHI Grant Numbers JP20H00354 and the CREST(JPMJCR15F3), JST. The research leading to these results has received funding from the European Union's Horizon 2020 research and innovation program under grant agreements no. 785219-Graphene Core2 and 881603-Graphene Core3.

**TOC Graphic**

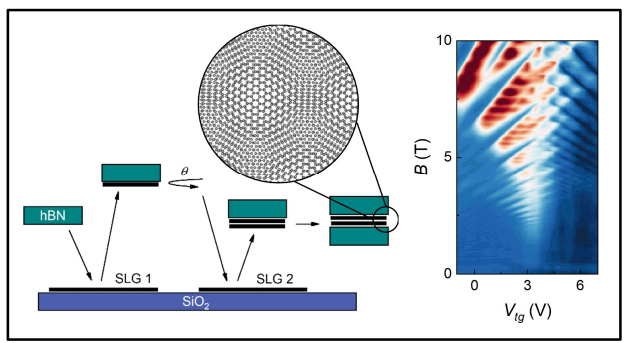